\documentclass[rnote]{aa}
\usepackage{graphicx}

\newcommand{\prp}    {${\rlap.}^{\prime}$}
\newcommand{\grp}    {${\rlap.}^{\circ}$}
\newcommand{\pri}    {${\rlap.}^{\prime \prime}$}
\newcommand{\rl}     {${\rlap.}^{s}$}

\newcommand{\ltsima} {$\; \buildrel < \over \sim \;$}
\newcommand{\simlt}  {\lower.5ex\hbox{\ltsima}}            % < over MMM
\newcommand{\gtsima} {$\; \buildrel > \over \sim \;$}
\newcommand{\simgt}  {\lower.5ex\hbox{\gtsima}}            % > over MMM

\begin{document}

\title{A radio and infrared exploration of the Cygnus X-3 environments}

\author{
J. Mart\'{\i}\inst{1,4}
\and D. P\'erez-Ram\'{\i}rez\inst{1,4}
\and P. Luque-Escamilla\inst{2,4}
\and J.~L. Garrido\inst{1,4}
\and J.~M. Paredes\inst{3}
\and A.~J. Mu\~noz-Arjonilla\inst{4}
\and J.~R. S\'anchez-Sutil\inst{4}
}

\offprints{J. Mart\'{\i}}

\institute{Departamento de F\'{\i}sica, EPS,  
Universidad de Ja\'en, Campus Las Lagunillas s/n, Edif. A3, 23071 Ja\'en, Spain \\
\email{jmarti@ujaen.es, dperez@ujaen.es, jlg@ujaen.es} 
\and Dpto. de Ing. Mec\'anica y Minera, EPS,
Universidad de Ja\'en, Campus Las Lagunillas s/n, Edif. A3, 23071 Ja\'en, Spain \\
\email{peter@ujaen.es}
\and Departament d'Astronomia i Meteorologia, 
Universitat de Barcelona, Av. Diagonal 647, 08028 Barcelona, Spain \\
\email{jmparedes@ub.edu}
\and
Grupo de Investigaci\'on FQM-322,
Universidad de Ja\'en, Campus Las Lagunillas s/n, Edif. A3, 23071 Ja\'en, Spain \\
%\email{}
}

\date{Received / Accepted}

\titlerunning{A radio and IR exploration of the Cyg X-3 environments}

\abstract
{}
{To confirm, or rule out, the possible hot spot nature of two
previously detected radio sources in the vicinity of the Cygnus X-3
microquasar.}
{We present the results of a radio and near infrared exploration of
the several arc-minute field around the well known galactic
relativistic jet source Cygnus X-3 using the Very Large Array and the
Calar Alto 3.5~m telescope.}
{The data this paper is based on do not presently support the hot spot
hypothesis.  Instead, our new observations suggest that these sources
are most likely background or foreground objects. Actually, none of
them appears to be even barely extended as would be expected if they
were part of a bow shock structure. Our near infrared observations
also include a search for extended emission in the Bracket $\gamma$
(2.166 $\mu$m) and $H_{2}$ (2.122 $\mu$m) lines as possible tracers of
shocked gas in the Cygnus X-3 surroundings. The results were similarly
negative and the corresponding upper limits are reported.}
% conclusions heading (optional)
{}
\keywords{
stars: individual: \object{Cygnus X-3} -- radio continuum: stars -- X-rays: binaries 
}

\maketitle

\section{Introduction} 

Cygnus X-3 is among the most intensively studied microquasars in the
Galaxy. The system is a high-mass X-ray binary with a WN Wolf-Rayet
companion star (see e.g. \cite{f99} and references therein) seen
through a high interstellar absorption ($A_V$\gtsima 10 mag) that
renders the optical counterpart undetectable in the visual domain.
Cygnus X-3 has been observed to undergo strong radio flares with flux
density increments by almost three orders or magnitude above the
normal quiescent level of $\sim0.1$ Jy at cm wavelengths.  The first
of them was the historic flare extensively described by \cite{g72} and
subsequent papers.  Relativistic jets from this microquasar have been
reported soon after some of these events outflowing collimately in the
North-South direction (see e.g. \cite{marti01}, \cite{miller04}). The
reader is referred to these papers for a more detailed account on the
flaring and sub-arcsecond properties of the source.

In recent times, concerns have arised about the possible effects of
continuate energy and momentum injection into the interstellar medium
(ISM) during the flaring lifetime of microquasar systems. These
effects appear to be far more important than previously thought. The
recent report of a `dark jet' in Cygnus X-1 by \cite{ga2005}
illustrates a likely case of a black hole microquasar silently
evacuating a significant fraction of its accretion power into its
surroundings and affecting them on a several pc scale.

In a previous paper, \cite{marti05} have reported the existence of two
possible hot spot candidates (HSCs) associated with Cygnus X-3, thus
suggesting an analogy with Fanaroff-Riley type II radio galaxies.
Hereafter and following their original designation, we will refer here
to these objects as the hot spot candidate North and South (HSCN and
HSCS), respectively.  They appeared as faint radio sources with
non-thermal spectrum at angular distances of 7\prp 07 and 4\prp 36
from Cygnus X-3. Moreover, the line joining them was within one degre
of the almost North-South position angle of the inner arc-second radio
jets as measured by \cite{marti01}.

The \cite{marti05} results, however, were hampered by the fact that
the HSCs were far from the phase centre of the interferometric array
and consequently suffered from significant bandwidth
smearing. Searches for near infrared counterparts provided also
negative results within the sensitivity of Two Micron All Sky Survey
(2MASS).

In this paper, we report new radio and near infrared observations of
both the HSCs and the Cygnus X-3 nearby environments. The improved
observational data do not confirm the proposed hot spot nature and
indicate that they are most likely background or foreground objects.
This fact leaves open again for Cygnus X-3 the issue of searching for
signatures of energy deposition from its relativistic jets into the
ISM. In the following sections we describe and discuss our recent
observational work that finally leads us to such different conclusion.

\section{Very Large Array radio observations}

Very Large Array (VLA) observations were carried out with the antennae
in B configuration on 2005 April 16 under clear sky conditions.  The
pointings were centered at the position of the northern and southern
HSCs as reported in \cite{marti05}. The targets were observed at both
the 6 cm and 3.5 cm wavelengths, equivalent to frequencies of 4.8 and
8.4 GHz. The instrumental setup included two intermediate frequency
(IF) pairs with 50 MHz bandwidth each.  The amplitude calibrator used
was 1331+305 (3C286) which is one of the two VLA primary
calibrators. The strong compact source 2007+404, within a few degrees
of Cygnus X-3, was observed as phase calibrator before and after each
science pointing and always for a couple of minutes at both
wavelengths.  Its corresponding bootstrapped flux density of 2007+404
was found to be $2.38 \pm 0.02$ Jy and $2.54 \pm 0.01$ Jy at 6 and 3.5
cm, respectively.  The data were edited processed using the AIPS
software package of NRAO following the standard procedures for
continuum calibration of interferometer data.  The final maps were
computed using the IMAGR task of AIPS based on the CLEAN deconvolution
algorithm.

%------------------------------------------------------------------------------
\begin{table}
%\begin{center}
\caption{Results of radio observations with the VLA.}
\label{vla}
\begin{tabular}{ccc}
\hline
HSC        &   $\alpha_{\rm J2000.0}$,    $\delta_{\rm J2000.0}$  &   Flux density      \\
           &                                                      &      (mJy)          \\
\hline
North      &  $20^h32^m$26\rl 874$\pm$0\rl 001                &  $S_{\rm 6~cm} = 2.01\pm 0.03$     \\
           &  $+41^{\circ}04^{\prime}$33\pri 02$\pm$0\pri 01  &  $S_{\rm 3.5~cm} = 1.33 \pm 0.03$  \\
           &                                                  &                                    \\               
South$^*$  &  $20^h 32^m$24\rl 97$\pm$0\rl 02                 &  $S_{\rm 6~cm} \leq 0.16^{**}$     \\
           &  $+40^{\circ} 53^{\prime}$05\pri 9$\pm$0\pri 2   &  $S_{\rm 3.5~cm} \leq 0.11^{**}$   \\
\hline
\end{tabular}
~\\
$^*$ Position taken from \cite{marti05}.\\
$^{**}$ Upper limits are given as $4 \sigma$ from VLA data of this paper.\\ 
%\end{center}
\end{table}
%------------------------------------------------------------------------------

%-------------------------------------------------------------
   \begin{figure}
   \centering
\resizebox{\hsize}{!}{\includegraphics{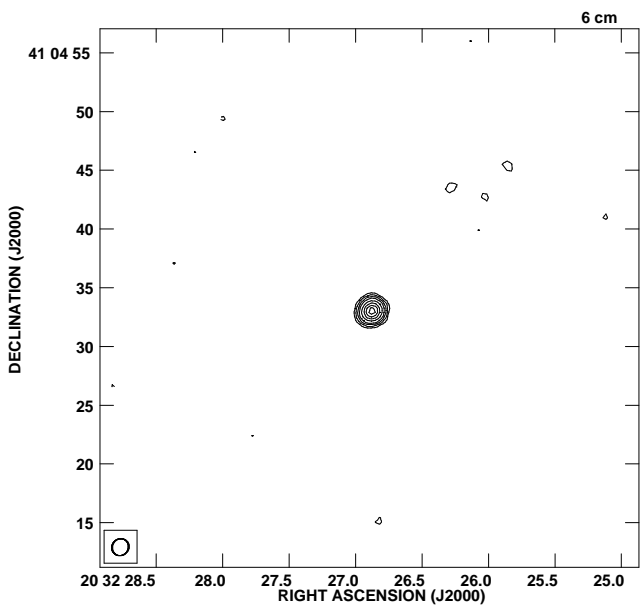}}
      \caption{VLA map at the 6 cm wavelength of the originally
proposed HSCN as observed on 2005 April 16.  The object is detected as
a clearly compact source. Contours are $-3$, 3, 5, 8, 12, 20, 30, 40
and 50 times 0.035 mJy beam$^{-1}$, the rms noise.  The ellipse shown
at the bottom left cornes is the synthesized beam equivalent to 1\pri
48$\times$1\pri 38, with position angle of $-$38\grp 2.  }
      \label{vla_hscn}
   \end{figure}
%-------------------------------------------------------------

%-------------------------------------------------------------
   \begin{figure*}
   \centering
\resizebox{\hsize}{!}{\includegraphics{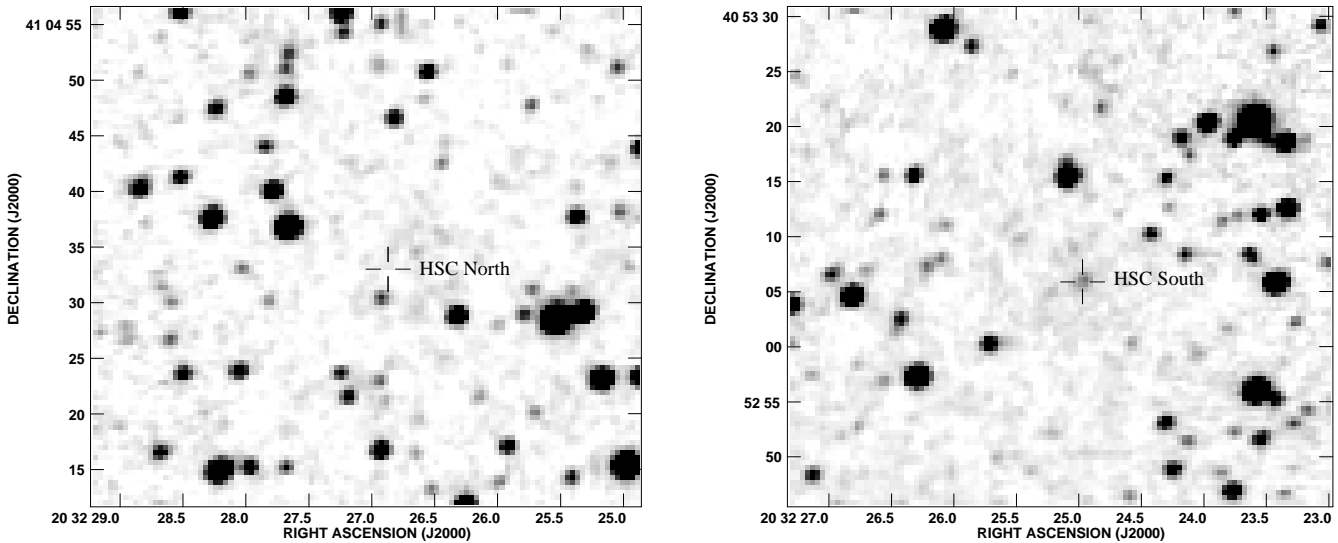}}
      \caption{The two fields of the previously proposed HSCs of
Cygnus X-3 as seen with the CAHA 3.5~m telescope plus the OMEGA2000
camera in the $Ks$-band. The crosses indicate the position of both the
HSCN and HSCS in the left and right pannel, respectively.  No near
infrared counterpart is detected for the northern object but a weak
source is clearly consistent with the southern one well within the
astrometric error.  }
      \label{caha_ns}
   \end{figure*}
%-------------------------------------------------------------

All radio information available is summarized in Table~\ref{vla}. Here
we can see how the HSCN was very well detected at both wavelengths.  A
contour map of our source is shown in Fig. \ref{vla_hscn} at the 6 cm
wavelength.  Based on the measured flux densities, the corresponding
spectral index is clearly non-thermal and can be estimated as $-0.74
\pm 0.05$. In contrast, only flux density upper limits were obtained
at the position of the HSCS.

\section{CAHA near infrared observations}

The positions of the HSCs were observed with the 3.5 m telescope at
the Centro Astron\'omico Hispano Alem\'an (CAHA) in Almer\'{\i}a
(Spain) on 2005 April 29th.  The OMEGA2000 camera was used and the
images were taken through the $J$, $H$, $Ks$, Br$\gamma$ and H$_2$
filters. This instrument consists of a Rockwell HAWAII2 HgCdTe
detector with $2048\times2048$ pixels sensitive within the wavelength
range from 0.8 to 2.5 $\mu$m.

Data reduction was carried out using the IRAF package. Firstly, the
sky was determined from the science frames themselves by
median-combining all of them for each band in order to obtained a
single image that was used for the sky background subtraction.
Secondly, dome flats were obtained for each band in series of frames
taken in pairs with lamps on and lamps off for eliminating the thermal
emission of the screen and dome surroundings. The actual flatfield is
then the difference image (lamp on - lamp off), thus allowing to
remove at the same time any dark count signal from the detector. A
flat field image for each band was then normalized and used for
correcting the rest of frames for each specific wave band.

After background subtraction and flatfielding, the data was
median-combined resulting in a deep image for each wave
band. Astrometry on the final frames was determined by identifying
about twenty stars in the field for which positions were retrieved
using the 2MASS catalog. The plate solutions were established by means
of the AIPS task XTRAN with a third order polinomial fit. The
residuals were typically of 0\pri 1 or less.

\begin{figure}
   \centering
\resizebox{\hsize}{!}{\includegraphics{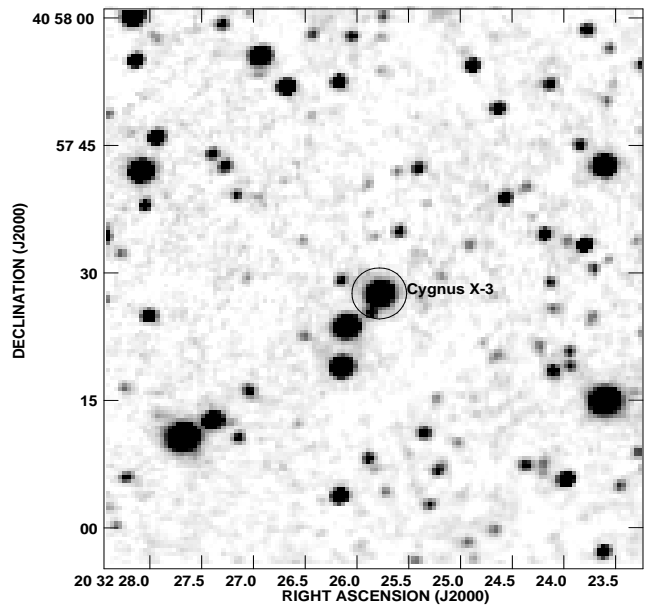}}
      \caption{CAHA 3.5~m $Ks$-band image of Cygnus X-3 taken on 2005
April 29 with OMEGA2000.  The well known near infrared counterpart of
this microquasar is located at the center of the frame and indicated
with a circle.  }
      \label{caha_ks}
   \end{figure}

The resulting images are shown in Figs. \ref{caha_ns} and
\ref{caha_ks} for both the HSCs and Cygnus X-3 itself in the $Ks$
band. They are significantly deeper than their 2MASS equivalents in
\cite{marti05}. The corresponding $Ks$ limiting magnitude is 18.7 mag,
while for $J$ and $H$-band images (not shown here) it is about 19.0
and 17.9, respectively.

%-------------------------------------------------------------
%-------------------------------------------------------------

Finally, we also produced narrow band images, namely in the Bracket
$\gamma$ (2.166 $\mu$m) and $H_{2}$ (2.122 $\mu$m) filters as possible
tracers of extended emission. Their continuum was subtracted using the
same $Ks$-band image appropriately scaled following the \cite{boker99}
procedure.  Nothing was detected at the HSC positions and upper limits
for any extended emission are given in the Discussion.

\begin{figure*}
   \centering
\resizebox{\hsize}{!}{\includegraphics{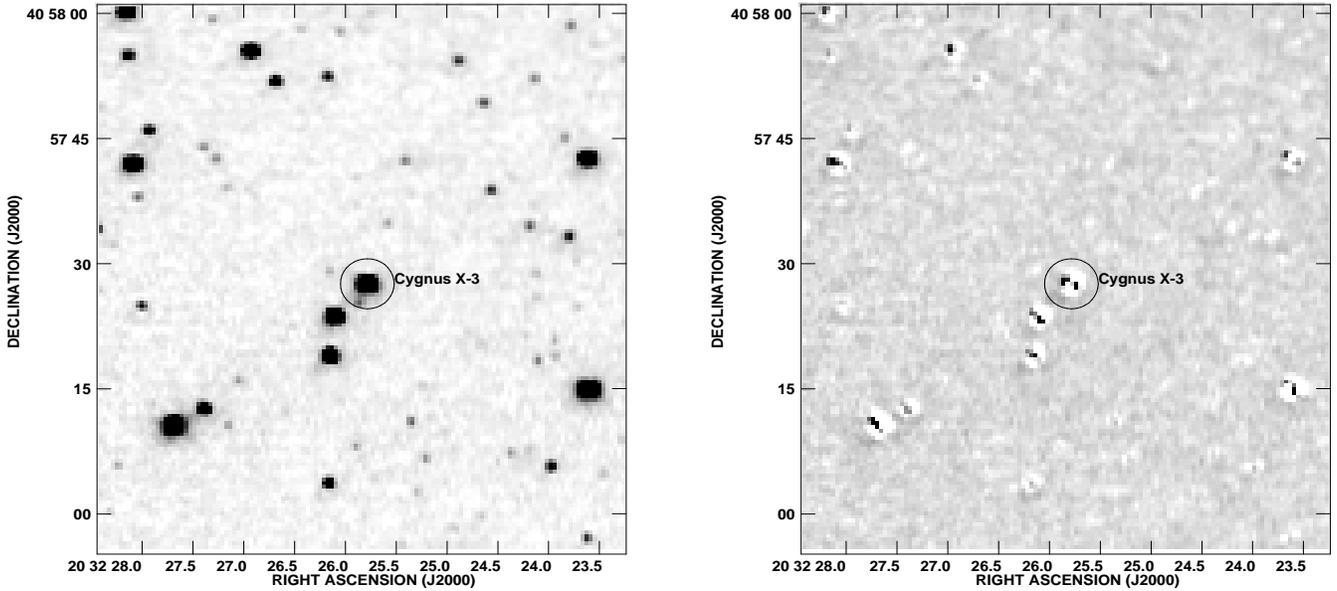}}
      \caption{{\bf Left.} CAHA 3.5~m image of Cygnus X-3 taken on
2005 April 29 with OMEGA2000 in the $H_2$ filter. {\bf Right.} The
same field after subtraction of the continuum in order to better
search for evidences of extended line emission. None was detected
within the sensitivity of our observations as it also occured with our
Br$\gamma$ narrow filter observations.  }
      \label{caha_h2}
   \end{figure*}

A representative sample of our line narrow band images in the close
vicinity of Cygnus X-3 are presented in Fig. \ref{caha_h2} for
illustration purposes, corresponding to the H$_2$ filter.

%-------------------------------------------------------------
%-------------------------------------------------------------

\section{Discussion}

\subsection{The possible hot spots in Cygnus X-3 revisited}

The observations presented in this paper were specially designed to
test the possible hot spot nature of the two radio sources reported in
\cite{marti05}. They were observed exactly at the phase center of the
VLA in order to avoid the smearing problems which affected the data in
this previous work. Nevertheless, the integration time devoted now to
the targets was half than in the former data.

As seen in Fig. \ref{vla_hscn}, the HSCN appears perfectly unresolved
at radio wavelengths with no trace of extended emission above a
$4\sigma$ upper limit of 0.14 mJy beam$^{-1}$.  Observing at 6 cm with
the VLA in B configuration would have allowed us to image radio
features with angular scales as large as $\sim36^{\prime \prime}$. The
fact that no extended radio emission in the several arc-second range
is detected above our sensitivity limit suggests, although does not
strictly rule out, that HSCN is likely not a hot spot.

In the HSCS case, we did not detect this source possibly because of
unsufficient integration time.  The upper limit in Table \ref{vla} is
roughly consistent with the flux density reported in the much more
sensitive but smearing-affected observations of \cite{marti05}.
Therefore, we cannot make any statement based on its apparent angular
size.

Our search for infrared counterparts with the 3.5~m CAHA telescope is
deeper than the 2MASS images in \cite{marti05} by more than four
magnitudes. The panels in Fig. \ref{caha_ns} clearly show that the
HSCN does not have an infrared counterpart within our improved
sensitivity limits in any of the filters. In contrast, the HSCS does
exhibit a faint infrared counterpart in almost perfect coincidence
with the VLA radio position originally reported in
\cite{marti05}. This object is detected only in the $Ks$ band.  The
offset between the VLA and CAHA positions is $\sim$0\pri 1 which is
well within the residual errors of the astrometric solution for the
OMEGA2000 frame based on 20 reference stars from the 2MASS survey.

The infrared counterpart of HSCS is clearly of stellar appearance and
it is most likely a galactic star or an extragalactic background
object. In any case, the fact of HSCS being a point-like source
inmediately argues against its originally proposed hot spot nature.

\subsection{The Cygnus X-3 environments in Br$\gamma$ and H$_2$ lines}

In our attempts to search for a signature of interaction between the
Cygnus X-3 relativistic outflows and the interstellar medium, we also
imaged the field in line narrow bands. Using the traditional tools of
star formation studies, the Br$\gamma$ and H$_2$ filters were chosen
as likely tracers of shocked gas.  The resulting images (see
Fig. \ref{caha_h2}) do not display evidence of extended emission
neither at the HSC positions nor in the vicinity of Cygnus X-3 itself.
The corresponding $4\sigma$ upper limit corresponds to about 17.6 mag
arcsec$^{-2}$ in both filters.

These results stress again the dichotomy between collimated outflows
from young stellar objects (YSOs) and microquasars concerning the
difficulty in detecting signatures of jet-environment
interaction. YSOs being born in dense molecular clouds comparatively
exhibit more often evidences of jet shocked gas around them
(\cite{he02}).

\subsection{The arc-second close vicinity of Cygnus X-3}

We also devoted our attention to the very close surroundings of Cygnus
X-3 in our deep CAHA $Ks$ infrared observations.  In particular, we
tried to look for indications of any elongation of Cygnus X-3 with
respect to the stellar profiles of the nearby stars.  It has been
reported by \cite{ogley97} that Cyg X-3 could be represented by two
stellar-type profiles 0\pri 56 apart in $K$-band, which may be
interpreted as an extended emission feature.  In order to characterize
the tridimensional profiles of Cyg X-3 and neighbour stars in the
field, we have obtained four bidimensional profiles in the main
directions for each of the sources by using the SLICE task in AIPS.
These profiles exhibits always a typical Gaussian behaviour and no
evidence of extended infrared emission have been found for Cyg
X-3. Given that the seeing in our images is estimated as 1\pri 07, we
should be able to separate only arc-second extended components if
present at the time of our observations and this was actually not the
case.  Nevertheless, this result does not exclude that a faint
elongation due to an infrared jet could be detected soon after a
strong outburst in analogy with the case of GRS1915+105 as observed by
\cite{sams96}.

\section{Conclusions}

After conducting intensive radio and near infrared observations of
Cygnus X-3 and its surroundings, our conclusions can be summarized as
follows:

\begin{enumerate}

\item The previously proposed hot spot nature of two nearby radio
sources in almost perfect alignment with the relativistic jets of this
microquasar has not been possible to be confirmed.  The absence of
clear indications for these objects being hot spots, such as extended
emission and no stellar-like counterpart for both of them, leads us to
be very cautious about our former hypothesis.

\item We have also presented deep line narrow band images of the
Cygnus X-3 field and provided strong upper limits for the brightness
of extended emission due to possible shocked gas.

\item The observational data collected and reported in the present
paper clearly show that searching for signatures of jet-ISM
interaction is a very difficult task which would require a more
sensitive sampling on a wide range of angular scales.

\item Finally, a comparison of the Cygnus X-3 stellar profile to that
of other stars in the field does not indicate evidences of additional
components within \gtsima1$^{\prime\prime}$ at near infrared
wavelegths.

\end{enumerate}

\begin{acknowledgements}
The authors acknowledge support by DGI of the Ministerio de
Educaci\'on y Ciencia (Spain) under grants AYA2004-07171-C02-02 and
AYA2004-07171-C02-01, FEDER funds and Plan Andaluz de Investigaci\'on
of Junta de Andaluc\'{\i}a as research group FQM322.  The National
Radio Astronomy Observatory is a facility of the National Science
Foundation operated under cooperative agreement by Associated
Universities, Inc. in the USA. This paper is also based on
observations collected at the Centro Astron\'omico Hispano Alem\'an
(CAHA) at Calar Alto, operated jointly by the Max-Planck Institut
f\"ur Astronomie and the Instituto de Astrof\'{\i}sica de
Andaluc\'{\i}a (CSIC).  This research has made use of the SIMBAD
database, operated at CDS, Strasbourg, France.  This publication makes
use of data products from the Two Micron All Sky Survey, which is a
joint project of the University of Massachusetts and the Infrared
Processing and Analysis Center/California Institute of Technology,
funded by the National Aeronautics and Space Administration and the
National Science Foundation in the USA.  The research of DPR has been
supported by the Education Council of Junta de Andaluc\'{\i}a (Spain).
\end{acknowledgements}

\end{document}